\documentclass[12pt]{article}

\usepackage{epsf}
\usepackage[dvips]{graphicx}

\usepackage{epic}
\usepackage{eepic}
\def\bc{\begin{center}}
\def\ec{\end{center}}
\def\beq{\begin{equation}}
\def\eeq{\end{equation}}

\def\hs#1{\hspace*{#1cm}}
\def\av#1{\langle #1 \rangle}

\begin{document}

\begin{center}
{\bfseries
SPACE-TIME PICTURE OF THE STRING FRAGMENTATION\\
AND
THE FUSION OF COLOUR STRINGS
}

\vskip 5mm

V.V. Vechernin

\vskip 5mm

{\small
{\it
St.-Petersburg State University
}
\\
{\it
E-mail: vechernin@pobox.spbu.ru
}}
\end{center}

\vskip 5mm

\begin{center}
\begin{minipage}{150mm}
\centerline{\bf Abstract}
It is shown that na\"{\i}ve two stage scenario of the soft multiparticle production in
hadronic and nuclear collisions at high energy,
when at first stage the colour strings are formed
and at the second stage these strings,
or some other (higher colour) strings
formed due to fusion of primary strings, are decaying,
emitting observed particles,
encounters
some difficulties at the attempt to analyse the space-time picture of the process.
Simple analysis shows the dominant is the process when the formation
and the decay of a string occur in parallel - a string breaks into two parts
already at rather small length (about 1$\div$2 fm in its c.m. system), then the process repeats
in the pieces and so on.
Nevertheless
it is proved to be possible to agree the string fusion idea with the space-time picture
of a string decay.
In the framework of the Artru-Mennessier model of a string fragmentation
the simple interpretation of the homogeneity of the rapidity distribution  for hadrons
produced from the decay of a single string at high energy is presented and the
analytical estimate for the density of this rapidity distribution is obtained.
\end{minipage}
\end{center}

\vskip 10mm

\def\figone{
\begin{figure}[t]
\bc
\unitlength=7mm
\begin{picture}(20,5)(-10,-0.5)
{\large

\linethickness{0.7 pt}   
\put(-10,0){\vector(1,0){20}}
\multiput(-3.6,0)(7.2,0){2}{\line(0,-1){0.2}}
\multiput(-7.7,0)(15.4,0){2}{\line(0,-1){0.2}}
\put(-3.6,-0.5){\makebox(0,0)[t]{$y^{}_2$}}
\put(3.6,-0.5){\makebox(0,0)[t]{$y^{}_3$}}
\put(-7.7,-0.5){\makebox(0,0)[t]{$y^{}_1$}}
\put(7.7,-0.5){\makebox(0,0)[t]{$y^{}_4$}}
\put(10,-0.5){\makebox(0,0)[t]{$y$}}

\multiput(-3,2.8)(2,0){4}{\vector(0,1){1}}
\multiput(-3,1.5)(2,0){4}{\vector(0,-1){1}}
\allinethickness{1 pt}   
\multiput(-6.9,2.8)(2,0){8}{\vector(1,4){.25}}
\multiput(-7.1,2.8)(2,0){8}{\vector(-1,4){.25}}
\multiput(-6.9,1.5)(2,0){8}{\vector(1,-4){.25}}
\multiput(-7.1,1.5)(2,0){8}{\vector(-1,-4){.25}}

\linethickness{2 pt}   
\allinethickness{2 pt}   

\put(-3.5,2){\oval(.5,.5)[l]}
\put(3.5,2){\oval(.5,.5)[r]}
\linethickness{1.5 pt}   
\put(-3.5,2.31){\line(1,0){7}}
\put(-3.5,1.81){\line(1,0){7}}

\put(-7.5,2.1){\oval(.5,.5)[l]}
\put(7.5,2.1){\oval(.5,.5)[r]}
\put(-7.5,2.41){\line(1,0){15}}
\put(-7.5,1.91){\line(1,0){15}}

}
\end{picture}
\ec
\caption[dummy]{\label{fig:overlap} \label{fig:fusion}
The string overlap in rapidity ($y$) in two stage scenario.
}
\end{figure}
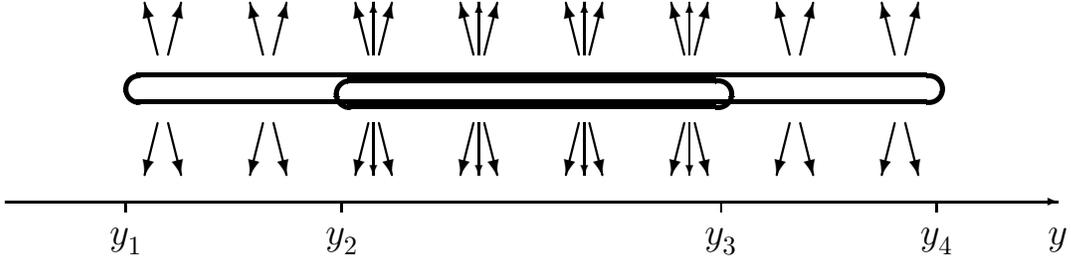
}
\def\figtwo{
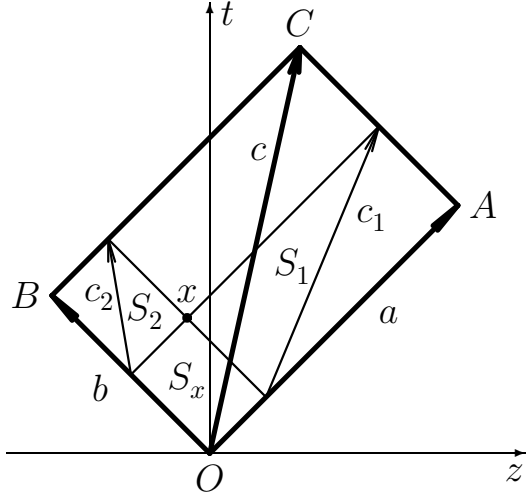
\begin{figure}[th]
\bc
\unitlength=3mm
\begin{picture}(20,20)(-5,0)
{\large

\linethickness{0.3 pt}   
\put(0,0){\vector(0,1){20}}
\put(-9,0){\vector(1,0){23}}

\allinethickness{2 pt}   
\drawline[100](0,0)(11,11)(4,18)(-7,7)(0,0)

\drawline[100](9.9,10.1)(11,11)
\drawline[100](10.1,9.9)(11,11)

\drawline[100](-6.1,5.9)(-7,7)
\drawline[100](-5.9,6.1)(-7,7)

\drawline[100](0,0)(4,18)
\drawline[100](3.55,16.9)(4,18)
\drawline[100](3.95,16.9)(4,18)

\put(-1,6){\circle*{0.2}}

\allinethickness{1 pt}   
\drawline[100](2.5,2.5)(-4.5,9.5)
\drawline[100](-3.5,3.5)(7.5,14.5)

\drawline[100](2.5,2.5)(7.5,14.5)
\drawline[100](6.9,13.5)(7.5,14.5)
\drawline[100](7.2,13.4)(7.5,14.5)

\drawline[100](-3.5,3.5)(-4.5,9.5)
\drawline[100](-4.5,8.5)(-4.5,9.5)
\drawline[100](-4.2,8.5)(-4.5,9.5)

\put(-1,6.7){\makebox(0,0)[b]{$x$}}
\put(-1,4){\makebox(0,0)[t]{$S^{}_x$}}

\put(7.5,6.5){\makebox(0,0)[lt]{$a$}}
\put(-4.5,3.5){\makebox(0,0)[rt]{$b$}}
\put(2.5,13){\makebox(0,0)[rb]{$c$}}
\put(6.5,11){\makebox(0,0)[lt]{$c^{}_1$}}
\put(-4.2,7.5){\makebox(0,0)[rt]{$c^{}_2$}}

\put(11.5,11){\makebox(0,0)[l]{$A$}}
\put(-7.5,7){\makebox(0,0)[r]{$B$}}
\put(0,-0.5){\makebox(0,0)[t]{$O$}}
\put(4,18.5){\makebox(0,0)[b]{$C$}}

\put(0.5,19.5){\makebox(0,0)[l]{$t$}}
\put(13.5,-0.5){\makebox(0,0)[t]{$z$}}

\put(4.5,7.5){\makebox(0,0)[rb]{$S^{}_1$}}
\put(-3.7,5.5){\makebox(0,0)[lb]{$S^{}_2$}}
}
\end{picture}
\ec
\caption[dummy]{\label{fig:decay}
The string decay in the AMOR model.
}
\end{figure}
}
\def\figthree{
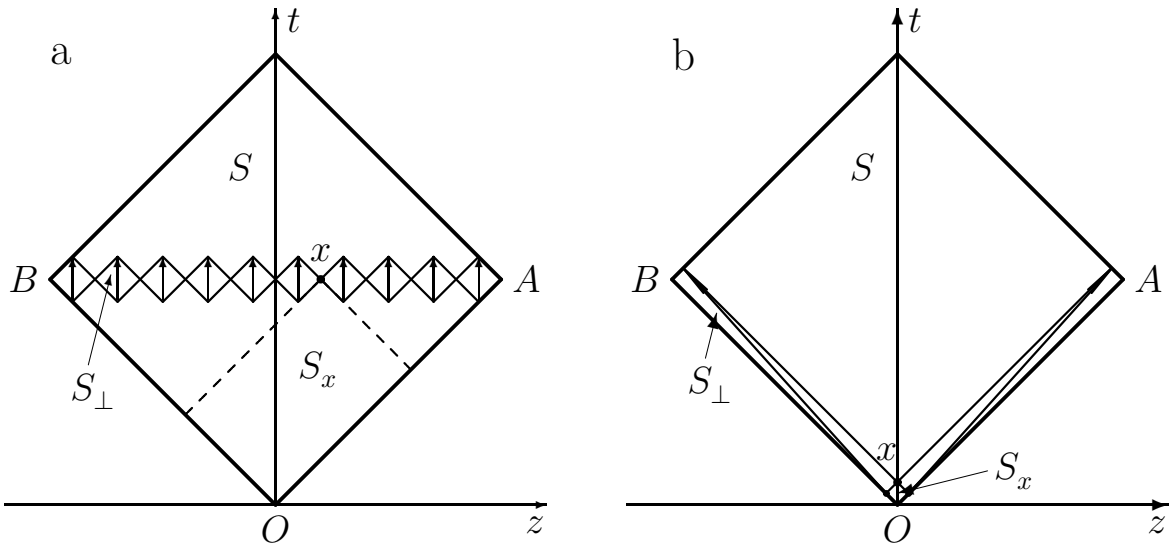
\begin{figure}[t]
\bc
\unitlength=3mm
{\large
\begin{picture}(20,23)(4,0)
\put(-10,20){\makebox(0,0)[l]{{\Large a}}}
\put(-8.3,6.2){\vector(1,4){1}}
\put(-8.1,5.8){\makebox(0,0)[t]{$S^{}_{\perp}$}}

\linethickness{0.5 pt}   
\put(0,0){\vector(0,1){22}}
\put(-12,0){\vector(1,0){24}}
\multiput(-9,9.1)(2,0){10}{\vector(0,1){1.9}}

\allinethickness{1 pt}   
\multiput(-9,9)(2,0){9}{\line(1,1){2}}
\multiput(-9,11)(2,0){9}{\line(1,-1){2}}

\allinethickness{1.5 pt}   
\drawline[100](0,0)(10,10)(0,20)(-10,10)(0,0)
\put(2,10){\circle*{0.2}}
\put(2,10.8){\makebox(0,0)[b]{$x$}}
\put(1,6){\makebox(0,0)[l]{$S^{}_x$}}
\put(-1,15){\makebox(0,0)[r]{$S$}}

\put(10.5,10){\makebox(0,0)[l]{$A$}}
\put(-10.5,10){\makebox(0,0)[r]{$B$}}
\put(0,-0.5){\makebox(0,0)[t]{$O$}}
\put(0.5,21.5){\makebox(0,0)[l]{$t$}}
\put(11.5,-0.5){\makebox(0,0)[t]{$z$}}

\allinethickness{0.8 pt}   
\dashline[60]{0.6}[0.05](-4,4)(2,10)
\dashline[60]{0.6}[0.05](6,6)(2,10)

\put(27,0){
\begin{picture}(20,23)(0,0)

\allinethickness{0.5 pt}   
\linethickness{0.5 pt}   
\put(-10,20){\makebox(0,0)[l]{{\Large b}}}
\put(-8.5,6.5){\vector(1,4){0.5}}
\put(4,1.5){\vector(-4,-1){4}}
\put(-8.3,6.1){\makebox(0,0)[t]{$S^{}_{\perp}$}}
\put(0,0){\vector(0,1){22}}
\put(-12,0){\vector(1,0){24}}

\allinethickness{1 pt}   
\put(-0.5,0.5){\line(1,1){10}}
\put(0.5,0.5){\line(-1,1){10}}

\allinethickness{1 pt}   

\put(0.5,0.5){\circle*{0.2}}
\drawline[100](0.5,0.5)(9.5,10.5)
\drawline[100](8.6,9.4)(9.5,10.5)
\drawline[100](8.45,9.55)(9.5,10.5)

\put(-0.5,0.5){\circle*{0.2}}
\drawline[100](-0.5,0.5)(-9.5,10.5)
\drawline[100](-8.6,9.4)(-9.5,10.5)
\drawline[100](-8.45,9.55)(-9.5,10.5)

\allinethickness{1.5 pt}   
\drawline[100](0,0)(10,10)(0,20)(-10,10)(0,0)
\put(0,1){\circle*{0.2}}
\put(0,2){\makebox(0,0)[br]{$x$}}
\put(4.3,1.5){\makebox(0,0)[l]{$S^{}_x$}}
\put(-1,15){\makebox(0,0)[r]{$S$}}

\put(10.5,10){\makebox(0,0)[l]{$A$}}
\put(-10.5,10){\makebox(0,0)[r]{$B$}}
\put(0,-0.5){\makebox(0,0)[t]{$O$}}
\put(0.5,21.5){\makebox(0,0)[l]{$t$}}
\put(11.5,-0.5){\makebox(0,0)[t]{$z$}}

\end{picture}
} 
\end{picture}
}
\ec
\caption[dummy]{\label{fig:rare}
Two examples of the rare string decays.
}
\end{figure}
}
\def\figfour{
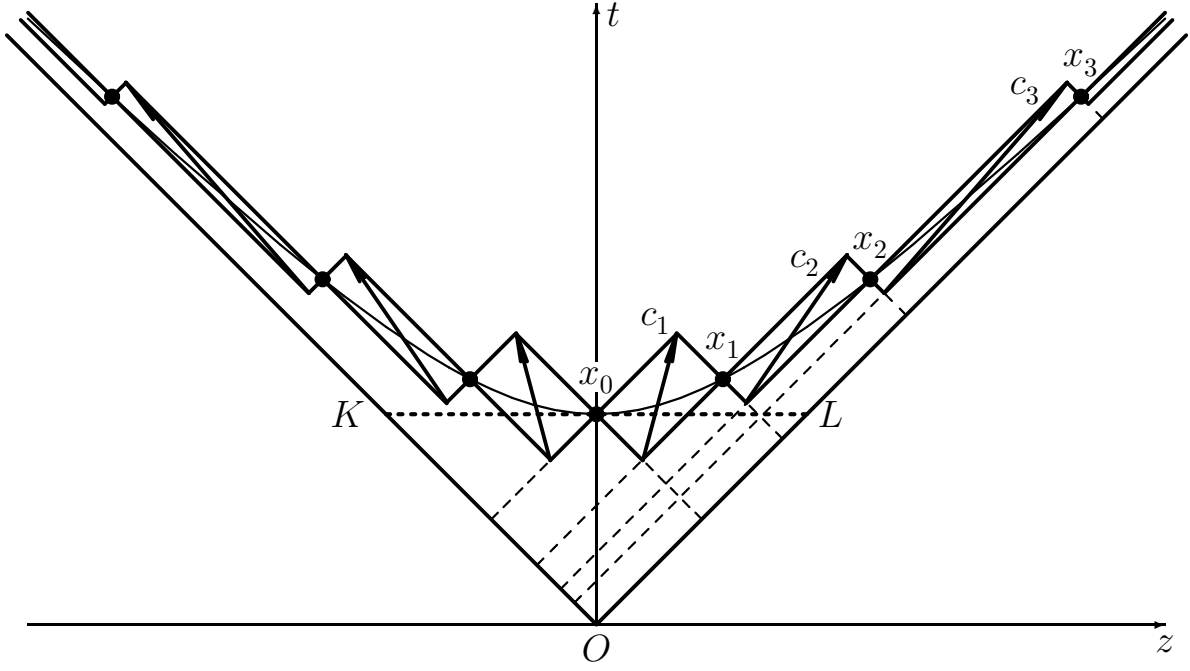
\begin{figure}[t]
\bc
\unitlength=2.8mm
{\large
\begin{picture}(54,30)(-27,0)

\linethickness{0.5 pt}   
\put(0,12.5){\vector(0,1){17}}
\put(0,0){\line(0,1){11}}
\put(-27,0){\vector(1,0){54}}

\allinethickness{1.5 pt}   

\put(0,10){\line(1,1){3.8}}  
\put(0,10){\line(1,-1){2.2}} 
\put(6,11.66){\line(-1,-1){3.8}}
\put(6,11.66){\line(-1,1){2.2}}

\put(0,10){\line(-1,1){3.8}}  
\put(0,10){\line(-1,-1){2.2}} 
\put(-6,11.66){\line(1,-1){3.8}}
\put(-6,11.66){\line(1,1){2.2}}

\put(6,11.66){\line(1,1){5.9}}
\put(6,11.66){\line(1,-1){1.1}}
\put(13,16.4){\line(-1,-1){5.9}}
\put(13,16.4){\line(-1,1){1.1}}

\put(-6,11.66){\line(-1,1){5.9}}
\put(-6,11.66){\line(-1,-1){1.1}}
\put(-13,16.4){\line(1,-1){5.9}}
\put(-13,16.4){\line(1,1){1.1}}

\put(13,16.4){\line(1,1){9.35}}
\put(13,16.4){\line(1,-1){0.65}}
\put(23,25.08){\line(-1,-1){9.35}}
\put(23,25.08){\line(-1,1){0.65}}

\put(-13,16.4){\line(-1,1){9.35}}
\put(-13,16.4){\line(-1,-1){0.65}}
\put(-23,25.08){\line(1,-1){9.35}}
\put(-23,25.08){\line(1,1){0.65}}

\put(23,25.08){\line(1,1){4}}
\put(23,25.08){\line(1,-1){0.35}}
\put(23.35,24.73){\line(1,1){4}}

\put(-23,25.08){\line(-1,1){4}}
\put(-23,25.08){\line(-1,-1){0.35}}
\put(-23.35,24.73){\line(-1,1){4}}

\allinethickness{1 pt}   

\drawline[100](0,10)(1,10.05)(2,10.2)(3,10.44)(4,10.77)(5,11.18)(6,11.66)(7,12.21)(8,12.81)(10,14.14)(12,15.62)(13,16.4)(15,18.03)(20,22.36)(23,25.08)(25,26.93)(27,28.79)
\drawline[100](0,10)(-1,10.05)(-2,10.2)(-3,10.44)(-4,10.77)(-5,11.18)(-6,11.66)(-7,12.21)(-8,12.81)(-10,14.14)(-12,15.62)(-13,16.4)(-15,18.03)(-20,22.36)(-23,25.08)(-25,26.93)(-27,28.79)

\allinethickness{1.5 pt}   
\drawline[100](-28,28)(0,0)(28,28)

\allinethickness{3 pt}   
\put(0,10){\circle*{0.4}}
\put(6,11.66){\circle*{0.4}}
\put(13,16.4){\circle*{0.4}}
\put(-6,11.66){\circle*{0.4}}
\put(-13,16.4){\circle*{0.4}}
\put(23,25.08){\circle*{0.4}}
\put(-23,25.08){\circle*{0.4}}
\put(0,11){\makebox(0,0)[b]{$x^{}_0$}}
\put(6,12.66){\makebox(0,0)[b]{$x^{}_1$}}
\put(13,17.4){\makebox(0,0)[b]{$x^{}_2$}}
\put(23,26.08){\makebox(0,0)[b]{$x^{}_3$}}


\put(10.5,10){\makebox(0,0)[l]{$L$}}
\put(-11,10){\makebox(0,0)[r]{$K$}}
\put(0,-0.5){\makebox(0,0)[t]{$O$}}
\put(0.5,29){\makebox(0,0)[l]{$t$}}
\put(27,-0.5){\makebox(0,0)[t]{$z$}}

\allinethickness{1.5 pt}   
\dashline[100]{1}[1](-10,10)(10,10)
\allinethickness{0.8 pt}   
\dashline[60]{0.6}[0.05](-5,5)(0,10)
\dashline[60]{0.6}[0.05](5,5)(0,10)
\dashline[60]{0.6}[0.05](-2.83,2.83)(6,11.66)
\dashline[60]{0.6}[0.05](8.83,8.83)(6,11.66)
\dashline[60]{0.6}[0.05](-1.7,1.7)(13,16.4)
\dashline[60]{0.6}[0.05](14.7,14.7)(13,16.4)
\dashline[60]{0.6}[0.05](-1.04,1.04)(23,25.08)
\dashline[60]{0.6}[0.05](24.04,24.04)(23,25.08)

\allinethickness{1.5 pt}   
\put(2.8,13.8){\makebox(0,0)[b]{$c^{}_1$}}
\drawline[100](2.2,7.8)(3.8,13.8)
\drawline[100](3.3,12.5)(3.8,13.8)
\drawline[100](3.6,12.5)(3.8,13.8)

\drawline[100](-2.2,7.8)(-3.8,13.8)
\drawline[100](-3.3,12.5)(-3.8,13.8)
\drawline[100](-3.6,12.5)(-3.8,13.8)

\put(9.9,16.56){\makebox(0,0)[b]{$c^{}_2$}}
\drawline[100](7.1,10.56)(11.9,17.56)
\drawline[100](11,16.5)(11.9,17.56)
\drawline[100](11.25,16.3)(11.9,17.56)

\drawline[100](-7.1,10.56)(-11.9,17.56)
\drawline[100](-11,16.5)(-11.9,17.56)
\drawline[100](-11.25,16.3)(-11.9,17.56)

\put(20.35,24.75){\makebox(0,0)[b]{$c^{}_3$}}
\drawline[100](13.65,15.75)(22.35,25.75)
\drawline[100](20.9,24.2)(22.35,25.75)
\drawline[100](21.05,24)(22.35,25.75)

\drawline[100](-13.65,15.75)(-22.35,25.75)
\drawline[100](-20.9,24.2)(-22.35,25.75)
\drawline[100](-21.05,24)(-22.35,25.75)

\end{picture}
}
\ec
\caption[dummy]{\label{fig:dominant}
The dominant string decay.
}
\end{figure}
}
\def\figfive{
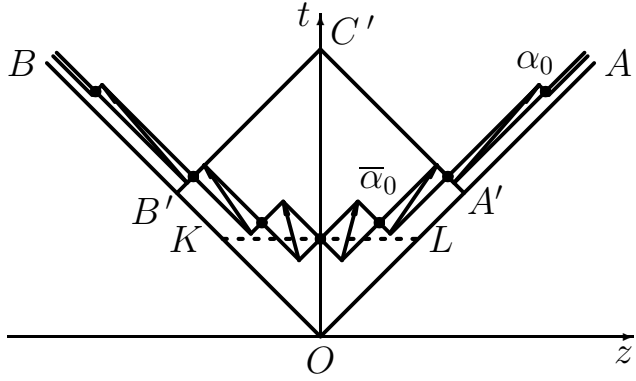
\begin{figure}[t]
\bc
\unitlength=1.3mm
{\large

\begin{picture}(64,33)(-32,0)

\linethickness{0.5 pt}   
\put(0,0){\vector(0,1){33}}
\put(0,0){\line(0,1){11}}
\put(-32,0){\vector(1,0){64}}

\allinethickness{1.5 pt}   

\put(0,10){\line(1,1){3.8}}  
\put(0,10){\line(1,-1){2.2}} 
\put(6,11.66){\line(-1,-1){3.8}}
\put(6,11.66){\line(-1,1){2.2}}

\put(0,10){\line(-1,1){3.8}}  
\put(0,10){\line(-1,-1){2.2}} 
\put(-6,11.66){\line(1,-1){3.8}}
\put(-6,11.66){\line(1,1){2.2}}

\put(6,11.66){\line(1,1){5.9}}
\put(6,11.66){\line(1,-1){1.1}}
\put(13,16.4){\line(-1,-1){5.9}}
\put(13,16.4){\line(-1,1){1.1}}

\put(-6,11.66){\line(-1,1){5.9}}
\put(-6,11.66){\line(-1,-1){1.1}}
\put(-13,16.4){\line(1,-1){5.9}}
\put(-13,16.4){\line(1,1){1.1}}

\put(13,16.4){\line(1,1){9.35}}
\put(13,16.4){\line(1,-1){0.65}}
\put(23,25.08){\line(-1,-1){9.35}}
\put(23,25.08){\line(-1,1){0.65}}

\put(-13,16.4){\line(-1,1){9.35}}
\put(-13,16.4){\line(-1,-1){0.65}}
\put(-23,25.08){\line(1,-1){9.35}}
\put(-23,25.08){\line(1,1){0.65}}

\put(23,25.08){\line(1,1){4}}
\put(23,25.08){\line(1,-1){0.35}}
\put(23.35,24.73){\line(1,1){4}}

\put(-23,25.08){\line(-1,1){4}}
\put(-23,25.08){\line(-1,-1){0.35}}
\put(-23.35,24.73){\line(-1,1){4}}

\allinethickness{1 pt}   

\allinethickness{1.5 pt}   
\drawline[100](-28,28)(0,0)(28,28)
\drawline[100](14.7,14.7)(0,29.4)
\drawline[100](-14.7,14.7)(0,29.4)

\allinethickness{3 pt}   
\put(0,10){\circle*{0.4}}
\put(6,11.66){\circle*{0.4}}
\put(13,16.4){\circle*{0.4}}
\put(-6,11.66){\circle*{0.4}}
\put(-13,16.4){\circle*{0.4}}
\put(23,25.08){\circle*{0.4}}
\put(-23,25.08){\circle*{0.4}}

\put(20,28){\makebox(0,0)[l]{$\alpha_0$}}
\put(4,16){\makebox(0,0)[l]{$\overline{\alpha}_0$}}
\put(15,14){\makebox(0,0)[l]{$A'$}}
\put(-15,13){\makebox(0,0)[r]{$B^{\,\prime}$}}
\put(29,28){\makebox(0,0)[l]{$A$}}
\put(-29,28){\makebox(0,0)[r]{$B$}}
\put(0,-1){\makebox(0,0)[t]{$O$}}
\put(1,30){\makebox(0,0)[lb]{$C^{\;\prime}$}}
\put(-1,33){\makebox(0,0)[r]{$t$}}
\put(31,-1){\makebox(0,0)[t]{$z$}}

\allinethickness{1.5 pt}   
\allinethickness{0.8 pt}   

\allinethickness{1.5 pt}   
\drawline[100](2.2,7.8)(3.8,13.8)
\drawline[100](3.3,12.5)(3.8,13.8)
\drawline[100](3.6,12.5)(3.8,13.8)

\drawline[100](-2.2,7.8)(-3.8,13.8)
\drawline[100](-3.3,12.5)(-3.8,13.8)
\drawline[100](-3.6,12.5)(-3.8,13.8)

\drawline[100](7.1,10.56)(11.9,17.56)
\drawline[100](11,16.5)(11.9,17.56)
\drawline[100](11.25,16.3)(11.9,17.56)

\drawline[100](-7.1,10.56)(-11.9,17.56)
\drawline[100](-11,16.5)(-11.9,17.56)
\drawline[100](-11.25,16.3)(-11.9,17.56)

\drawline[100](13.65,15.75)(22.35,25.75)
\drawline[100](20.9,24.2)(22.35,25.75)
\drawline[100](21.05,24)(22.35,25.75)

\drawline[100](-13.65,15.75)(-22.35,25.75)
\drawline[100](-20.9,24.2)(-22.35,25.75)
\drawline[100](-21.05,24)(-22.35,25.75)

\put(11,10){\makebox(0,0)[l]{$L$}}
\put(-12,10){\makebox(0,0)[r]{$K$}}
\allinethickness{1.5 pt}   
\dashline[300]{1}[1](-10,10)(10,10)

\end{picture}
}
\ec
\caption[dummy]{\label{fig:strfusion}
The space-time picture of the string fusion.
}
\end{figure}
}

\section{Introduction. AMOR model of string fragmentation}

Soft and semihard parts of the multiparticle production at high energy
are successfully described in terms of colour strings stretched between
the projectile and target \cite{Kaidalov,Capella1}
in the framework of a two-stage scenario, when
at the first stage a certain number of colour
strings stretched between the incoming partons are formed and at the second stage these
strings decay into the observed secondary hadrons.
In the case of nuclear collision, the number of strings grows
with the growing energy and atomic numbers of colliding nuclei, and
one has to take into account the interaction between
strings in the form of their fusion and/or percolation
\cite{BP1}-\cite{YF07} (see Fig.\ref{fig:overlap}).
The aim of the present paper is to analyse to what extent
the two-stage scenario is compatible
with the space-time picture of the process.

We'll consider the space-time evolution of a string in the framework of the classical
approach with the action
\beq
I=-\gamma\int\sqrt{(\dot{x} x')^2 - \dot{x}^2 x'^2}\  d\sigma d\tau \ .
\label{I}
\eeq
We'll also restrict our consideration to the simplest case of so-called "yo-yo" string.
As is well known (see, for example, \cite{BarbNest,Werner}) in this case
the motion of the string is the oscillations, the half-cycle of which is shown
as a rectangle $OACB$ on the space-time diagram in Fig.\ref{fig:decay}.

\figone

\figtwo

The 4-momentum $P$ of the string is connected
with the diagonal vector $c$ of this rectangle:
$P=\gamma c$ and the mass of the string $M$
is given by $M^2=P^2=\gamma^2 c^2$. One can decomposes diagonal vector $c$
on the sum of two light cone vectors: $c=a+b$; $a^2=b^2=0$, $a_-=b_+=0$, where
$a_\pm\equiv a_0\pm a_z$ and $b_\pm\equiv b_0\pm b_z$.
Then we have $M^2=2\gamma^2(ab)=\gamma^2a_+b_-=2\gamma^2|OA||OB|=2\gamma^2S_E(OACB)$,
as $a_+=a_0+a_z=2a_0=\sqrt{2}|OA|$ and $b_-=b_0-b_z=2b_0=\sqrt{2}|OB|$.
Here $S_E(OACB)$ is the Euclidean area of the rectangle $OACB$.

At first sight it seems a little bit strange that the squared mass of the string
is proportional to Euclidean area of the $OACB$, because the meaning
of the action $I$ is the area sweeping by the string in Minkowski space.
To clarify this point note, that the element of the area,
formed by two arbitrary vectors $a=(a_0,\bf{a})$ and $b=(b_0,\bf{b})$
in Minkowski space  $S_M$ is given by
$$
S^2_M=(ab)^2-a^2b^2=S^2-\Delta \ ,
$$
where $(ab)=(a_0b_0-\bf{a}\bf{b})$, $a^2=a_0^2-\bf{a}^2$, $b^2=b_0^2-\bf{b}^2$
and  $S^2=(a_0 {\bf b} - b_0 {\bf a})^2$, $\Delta=\bf{a}^2\bf{b}^2-(\bf{a}\bf{b})^2$.
The element of the area in Euclidean space $S_E$ is given by
$$
S^2_E=(a\times b)^2=a^2b^2-(ab)^2=S^2+\Delta \ ,
$$
where now $(ab)=(a_0b_0+\bf{a}\bf{b})$, $a^2=a_0^2+\bf{a}^2$, $b^2=b_0^2+\bf{b}^2$ and the
$S^2$ and the $\Delta$ are the same.
In the case of yo-yo string $\Delta=a_z^2 b_z^2-(a_z b_z)^2=0$ and hence $S_M=S_E=S$.
So for yo-yo string we have
\beq
P=\gamma c \ , \hs 1  M^2= 2\gamma^2 S=2\gamma^2S^{}_E(OACB)
\label{Pc}
\eeq

After the split of the string in the space-time point $x$ two strings with the momenta
$P_1=\gamma c_1$, $P_2=\gamma c_2$ and the masses
$M_1^2= 2\gamma^2 S_1$, $M_2^2= 2\gamma^2 S_2$  are formed
(see Fig.\ref{fig:decay}).
In principle any chain of rectangles connected by the corners
and going from the point $A$ to $B$ corresponds to some possible decay
of the initial string to substrings. At that the substrings with small area,
of order of particle masses, are associated with produced particles.

\figthree

In Fig.\ref{fig:rare} we present two specific examples of the string decay.
In Fig.\ref{fig:rare}a all decay points $x$ are at the same time,
when the length of the string is maximal in its c.m. system.
In this case all rapidities
of produced particles are equal to zero.
In Fig.\ref{fig:rare}b the decay point $x$ is very close to the origin $O$, so that
the area of two formed rectangles are of order of particle masses. In this case
the string decays on two particles with the minimal and maximal possible values of rapidity.
Both cases do not correspond to the typical physical situation, when one has more or less
homogeneous distribution of produced particles in rapidity. The reason is the small probability
of events in Fig.\ref{fig:rare}. The dominant process is shown in Fig.\ref{fig:dominant}  (see below).

We'll consider the string fragmentation in the framework of
the Artru-Mennessier AMOR model \cite{AM,A},
which is used in the VENUS event generator \cite{Werner}
and has in our opinion more fundamental physical foundations,
than the Lund model \cite{Lund}, used for example in PYTHIA
(see discussion in \cite{Werner,Lund}).
In the AMOR model the probability of the string split in the space-time point $x$
is proportional to the probability of the absence of splitting points in the area $S_x$
(see Fig.\ref{fig:decay}):
\beq
dP(x)=S^{-1}_0\, [1-P(x)]\, dS_x \ ,
\label{P(x)}
\eeq
which (by analogy with an unstable particle decay) leads to
\beq
P(x)=1-\exp{(-S_x/S_0)}\ , \hs1
dP(x)=S^{-1}_0\, \exp{(-S_x/S_0)}\, dS_x \ , \hs1
\av{S_x}=S_0  \ .
\label{dP(x)}
\eeq

\section {Analytical estimate of the rapidity distribution}
Let us now estimate the value of involving parameters.
The the string tension parameter $\gamma$ in (\ref{I}) is
connected with the slope $\alpha'$ of Regge trajectories: $\gamma^{-1}=2\pi \alpha'$ \cite{BarbNest}.
For $\alpha'$=0.9 GeV$^{-2}$ we have $\gamma$=0.18 GeV$^{2}$ ($c$=$\hbar$=1).
From the parameters of the potential connecting heavy quarks in nonrelativistic
models one obtains the close value $\gamma$=0.19 GeV$^{2}$. So we take
\beq
\gamma=0.18\, GeV^{2}=4.6 fm^{-2}=(0.47 fm)^{-2} \ .
\label{gamma}
\eeq

The parameter $S_0$, specifying the string decay probability in the AMOR model,
can be expressed through the dimensionless so-called 'area law parameter' $\alpha_0$
of the VENUS event generator:
$S^{-1}_0=2\alpha_0\gamma$. From the comparison of the VENUS event generator results
with the experimental data one finds in \cite{Werner}  $\alpha_0$=0.6. So we have
\beq
S_0=\av{S_x}=(2\alpha_0\gamma)^{-1}=4.6\, GeV^{-2}=0.18 fm^{2}=(0.43 fm)^{2} \ .
\label{S_0}
\eeq

\figfour

We introduce also the parameter $S_{0\perp}$=$\av{S_{\perp}}$ - the
mean area at which the string is associated with a produced particle
(see Fig.\ref{fig:rare}).
By (\ref{Pc}) $S_{0\perp}= m^2_{0\perp}/(2\gamma^2)$, where
$m^2_{0\perp}\equiv \langle m^2+p^2_{\perp}\rangle$ -
the average transverse mass of produced particles.
In this way one can effectively take into account the transverse momentum
of produced particles in the framework of yo-yo string model. So we have
\beq
S_{0\perp}=\av{S_{\perp}}=m^2_{0\perp}/(2\gamma^2)=\langle m^2+p^2_{\perp}\rangle/(2\gamma^2) \ .
\label{Sperp}
\eeq
The typical values of parameters $m^2_{0\perp}$ and $S_{0\perp}$
for different particles are presented below in the Table.
\bc
\begin{tabular}{c|c|c|c|c}
   &$m^2_{0\perp}$, GeV$^2$&$S_{0\perp}$, fm$^2$&$\beta$&$dN/dy$          \\ \hline
    $\pi$      &  0.11   &  0.07      &    0.4       &   1.5              \\ 
    $\rho$     &  0.6    &  0.36     &     2.0      &   0.75              \\ 
     $N$     &     1.0   &  0.6   &     3.3      &     0.63               \\ 
\end{tabular}
\ec

From the (\ref{S_0}) and the Table we see that values of $S_0$ and $S_{0\perp}$
are of the same order of magnitude: $S_0=(0.43 fm)^{2}$ and $S_{0\perp}=(0.26\div 0.78 fm)^{2}$.
At that the total area $S$ of the rectangle $OACB$, corresponding to the initial string
(see Fig.\ref{fig:rare}),  is much larger at high energies ($S\gg S_0,\ S_{0\perp}$).
For example, for a string with the mass $M^2=(100\,GeV)^2$
we find $S=(78 fm)^{2}$ and $|AB|=110 fm$.

Recalling formula (\ref{dP(x)}) we understand now the reason
of the small probability of the process in Fig.\ref{fig:rare}a.
In this case $S_x\sim S \gg S_0=\av{S_x}$ and the exponent in (\ref{dP(x)}) is small.
For the process in Fig.\ref{fig:rare}b we have
$S_x\ll S_{\perp}\sim S_{0\perp}\sim S_{0}=\av{S_x}\ll S$
and the probability of this process is small due to small phase volume
($S_x\ll \av{S_x}$, see (\ref{dP(x)})).
Clear that the dominant processes will be the ones with  $S_x\sim \av{S_x}=S_{0}$
(see Fig.\ref{fig:dominant}).

For rough estimate of the rapidity distribution
of produced particles in this case we'll consider that $S_x$ is equal for all splitting points:
$S_{x^{}_i}=S_{0}$ (see Figs.\ref{fig:rare} and \ref{fig:dominant}).
This leads to the condition
\beq
2S_{x^{}_i}=x^{2}_i=t^{2}_i-z^{2}_i=2S_{0}\ ,
\label{hyper}
\eeq
which means that all string splitting points $x^{}_i=(t^{}_i, z^{}_i)$
are situated on the hyperbola (\ref{hyper}).

We'll also suppose for rough estimate that the transverse masses of all produced particles is also equal.
By (\ref{Pc}) this leads to the condition $S_{i\perp}=S_{0\perp}$, which gives
(see Figs.\ref{fig:rare} and \ref{fig:dominant}):
\beq
2S_{i\perp}=c^2_i=2S_{0\perp}=m^2_{0\perp}/\gamma^2
  \ .
\label{Siperp}
\eeq

\figfive

For estimate
we'll also consider that the first split of a string occurs
in its middle,
as this situation of the point $x_0=(t_0,z_0)$ on the segment $KL$
corresponds to the maximal value of $S_{x^{}_0}$ (see Fig.\ref{fig:dominant}).
Note that the length of a string (in its c.m. system)
at the moment of the first split ($t_0$) is equal to
\beq
|KL|=2z_0=2\sqrt{2S_0}=1.2\, fm \ .
\label{split}
\eeq

After that the condition (\ref{Siperp}) fixes uniquely
the positions of all break points $x_i=(t_i,z_i)$ on the hyperbola (\ref{hyper}).
Then one can calculate all diagonal vectors $c_i$ in Fig.\ref{fig:dominant}
and find by (\ref{Pc}) the momenta
($p_i=\gamma c_i$) and rapidities ($y_i$)
of the produced particles. As a result one finds
\beq
y_i=(1/2)\ln(p_{i+}/p_{i+})=(1/2)\ln(c_{i+}/c_{i+})=(i-1/2)F(\beta) \ ,
\label{y_i}
\eeq
\beq
F(\beta)=\ln\, [\ 1+\beta/2+\sqrt{\beta(1+\beta/4)}\ ]\ \ ,
\label{F}
\eeq
\beq
\beta=S_{0\perp}/S_0=\alpha_0 m^2_{0\perp}/\gamma
=\alpha_0 \langle m^2+p^2_{\perp}\rangle/\gamma \ ,
\label{beta}
\eeq
where we used (\ref{S_0}) and (\ref{Sperp}). From (\ref{y_i}) we see that
the produced particles are homogeneously distributed in rapidity
and the density of this rapidity distribution is equal to
\beq
dN/dy=1/F(\beta)\ .
\label{dN}
\eeq

The numerical estimates by formulae (\ref{F})-(\ref{dN}) are presented above
for different values of the transverse mass parameter $m^2_{0\perp}$
in two last columns of the Table. One can see that we obtain the reasonable values
for such rough estimate. Note that we have supposed that the particles
only of one sort can be produced. This leads to 1.5 particles per unit of rapidity for pions
(including charged and neutral pions),
0.75 for $\rho$ mesons (which gives again 1.5 for pions) and 0.63 for nucleons.
In reality, when different particles can be produced in the decay of the string,
the reasonable value of charged particles produced from the decay of one string
per unit of rapidity is about 1.1 (see, for example, estimates in \cite{YF07}).

\section{Conclusion}

In conclusion we would like to discuss: is the string fusion picture shown in
Fig.\ref{fig:fusion} compatible with the dominant space-time picture of string fragmentation
shown in Fig.\ref{fig:dominant}? The answer is positive. One must only always keep in mind
that the picture of the string fusion in Fig.\ref{fig:fusion}
concerns the lengths of strings in rapidity space.
The corresponding space-time picture are shown in Fig.\ref{fig:strfusion}.
In this figure overlapping of two strings $OA'C'B'$ and $OACB$ with different masses
(different lengths in rapidity) is shown.

To take into account the string fusion
one has to use in the region $OA'C'B'$ the higher value $\overline{\alpha}_0$
of the area law parameter (\ref{S_0}),
describing the string fragmentation process, and
the usual value $\alpha_0$
outside this domain, in the rest of the region $OACB$.
From formulae (\ref{F})-(\ref{dN}) we see that the higher value of $\alpha_0$
(the lower value of $S_0$ (\ref{S_0}))
leads to the lower density of the produced particles rapidity distribution.

We see also in Fig.\ref{fig:strfusion} that
the formation
and the decay of the fused string occur in parallel and the string breaks into two parts
already at rather small length ($|KL|$ is about 1$\div$2 fm in the string c.m. system). This is compatible with the string fusion picture in rapidity space
shown in Fig.\ref{fig:fusion},
as by (\ref{y_i}) the particles with rapidities in the region $(y_2,y_3)$ in Fig.\ref{fig:fusion}
are being produced from the domain $OA'C'B'$ in Fig.\ref{fig:strfusion} with
higher value of the area law parameter $\alpha_0$.

The work was supported
by the RFFI grants 09-02-01327-a and 08-02-91004-CERN-a.


\begin{thebibliography}{99}
\bibitem{Kaidalov}
A.B.~Kaidalov, Phys. Lett. {\bfseries 116B}, 459 (1982);\\
A.B.~Kaidalov, K.A.~Ter-Martirosyan, Phys. Lett. {\bfseries 117B}, 247 (1982).
\bibitem{Capella1}
A.~Capella, U.P.~Sukhatme, C.--I.~Tan and J.~Tran Thanh Van,\\
Phys. Lett. {\bfseries B81}, 68 (1979); Phys. Rep. {\bfseries 236}, 225 (1994).
\bibitem{BP1}
M.A.~Braun, C.~Pajares, Phys. Lett. {\bfseries B287}, 154 (1992);
Nucl. Phys. {\bfseries B390}, 542 (1993).
\bibitem{PSM0}
N.S.~Amelin, M.A.~Braun, C.~Pajares, Phys. Lett. {\bfseries B306}, 312 (1993);\\
Z. Phys. {\bfseries C63}, 507 (1994).
\bibitem{PRL96}
N.~Armesto, M.A.~Braun, E.G.~Ferreiro, C.~Pajares,
Phys. Rev. Lett. {\bfseries 77}, 3736 (1996).
\bibitem{BPV00}
M.A.~Braun, C.~Pajares and V.V.~Vechernin, Phys. Lett. {\bfseries B493},  54 (2000).
\bibitem{EPJC04}
M.A.~Braun, R.S.~Kolevatov, C.~Pajares, V.V.~Vechernin,
Eur. Phys. J. {\bfseries '32}, 535 (2004).
\bibitem{YF07}
V.V.~Vechernin, R.S.~Kolevatov,
Physics of Atomic Nuclei, {\bfseries 70}, 1797; 1809 (2007).
\bibitem{BarbNest}
B.M.~Barbashov, V.V.~Nesterenko,
Relativistic String Model in Physics of Hadrons, \\ M., 1987, 176p. (in Russian).
\bibitem{Werner}
K.~Werner, Phys. Rep. {\bf 232}, 87 (1993).
\bibitem{AM}
X.~Artru, G.~Mennessier, Nucl. Phys. {\bf B70}, 93 (1974).
\bibitem{A}
X.~Artru, Phys. Rep. {\bf 97}, 147 (1983).
\bibitem{Lund}
B.~Andersson, G.~Gustafson, G.~Ingelman, T.~Sj\"{o}strand, Phys. Rep. {\bf 97}, 31 (1983).

\end{thebibliography}
\end{document}